\documentclass[a4paper]{llncs}

\usepackage{url}
\usepackage{graphicx}
\usepackage{tikz}
\usepackage{tkz-berge}

\usepackage{amssymb}
\usepackage{amsmath}
\usepackage{import}
\usepackage{amsmath}
\usepackage{xcolor}
\usepackage{xspace}
\usepackage{datatool}
\usepackage{glossaries}

\newcommand\setl[1]{\left\lbrace #1 \right\rbrace}

\newcommand{\tild}{\mathord{\sim}}

\newcommand{\F}{\mathcal{F}}
\newcommand{\jf}{{\m{\mathcal{J}\kern-0.2em\mathcal{F}}}}

\newcommand{\js}{{\mathcal{J}\kern-0.2em\mathcal{S}}}

\newcommand{\Edg}{\mathrm{Edge}}
\newcommand{\Path}{\mathrm{Path}}

\providecommand\m[1]{\ensuremath{#1}\xspace}
\renewcommand{\m}[1]{\ensuremath{#1}\xspace}
\newcommand{\trval}[1]{\m{\mathbf{#1}}}
\newcommand{\ltrue}{\trval{t}}
\newcommand{\lfalse}{\trval{f}}
\newcommand{\lunkn}{\trval{u}}
\newcommand{\lincon}{\trval{i}}
\newcommand{\Tr}{\ltrue}
\newcommand{\Fa}{\lfalse}
\newcommand{\Un}{\lunkn}
\newcommand{\In}{\lincon}
\newcommand{\rules}{\m{R}}

\newcommand{\leqk}{\m{\leq_k}}

\newcommand{\leqt}{\m{\leq_t}}

\begin{document}

\title{Extensions to Justification Theory}

\author{Simon Marynissen\inst{1,2}\orcidID{0000-0002-2894-1377}}

\authorrunning{S. Marynissen}

\institute{
  KU Leuven, Leuven, Belgium \and
  Vrije Universiteit Brussel, Brussels, Belgium
}
\maketitle

\section{Background}

Justification theory \cite{lpnmr/DeneckerBS15} is a unifying framework for semantics of  non-monotonic logics.
It is built on the notion of a justification, which intuitively is a graph that explains the truth value of certain facts in a structure.
Knowledge representation languages covered by justification theory include logic programs, argumentation frameworks, inductive definitions, and nested inductive and coinductive definitions.
In addition, justifications are also used for implementation purposes.
They are used to compute unfounded sets in modern ASP solvers \cite{ai/GebserKS12,ictai/DeCat13}, can be used to check for relevance of atoms in complete search algorithms \cite{ijcai/JansenBDJD16}, and recent lazy grounding algorithms are built on top of them \cite{jair/CatDBS15,ijcai/BogaertsW18}.

Semantics in justification theory are determined by \emph{branch evaluations}.
These are functions that map chains of facts to a single fact, in most case true, false or unknown.
For instance, in well-founded semantics, a positive loop is not allowed in a model, and thus the branch evaluation corresponding to the stable semantics maps positive loops to false.

A prototypical domain to apply knowledge representation and reasoning in is legal reasoning. I am particularly interested in an interactive decision enactment system, which makes legal decisions -- such as determining how many taxes someone should pay -- with interaction from the user. This interaction is in the form of the user supplying partial information.

In legal systems, a correct conclusion without a justification is frequently useless as you want to guarantee compliance with the law. My research will mainly be focussing on legal decisions that do not leave too much room for interpretations. Various types of public administration fall into this category.

In order to build such a system, a logic programming language with legal constructs is needed. Justification theory should then be used to define semantics for it since solutions of such a logic program need to be explained for it to be legally useful. The choice for justification theory is due to the fact that explanations are in the core of justification theory. However, many different approaches exist to provide explanations in answer set programming, see \cite{tplp/FandinnoS19} for a survey, but justification theory additionally provides a unifying framework for different semantics.

This motivates why we need to integrate deontic (prohibited, obligatory, or permitted) and alethic (possibility, impossibility and necessity) modal operators into justification theory since they are key constructs in legal texts. This language, and paired with that, extensions to justification theory will be the main driving factor throughout this extended abstract.

\subsection{Crash course in justification theory}

Before we dive into the actual material, a small crash course in justification theory is given. More detailed information can be found in \cite{nmr/MarynissenPBD18} and \cite{lpnmr/DeneckerBS15}.

Let $\F$ be any set of facts such that $\Tr, \Fa, \Un$ and $\In$ are in $\F$. They are the logical truth values \emph{true}, \emph{false}, \emph{unknown}, and \emph{inconsistent}\footnote{They have the same structure as the four-valued logic of Belnap \cite{Belnap77} with truth order $\Fa \leqt \Un, \In \leqt \Tr$ and information order $\Un \leqk \Fa,\Tr \leqk \In$.}. There is a negation $\tild$ on $\F$ such that $\tild x \neq x$ for non logical facts $x$ in $\F$ and $\tild(\tild x) =x$ for all facts $x$. The negations of the logical facts are $\tild \Tr=\Fa$, $\tild \Un = \Un$ and $\tild \In = \In$.
A \emph{justification frame} is a fact space $\F$ equipped with a set $\rules$ with elements of the form $x \gets A$ where $x$ is a non logical fact and $A$ is a nonempty subset of $\F$. The elements of $\rules$ are called \emph{rules} with \emph{head} $x$ and \emph{body} $A$. Each fact that occurs in the head of a rule is a \emph{defined} fact. All other facts are called \emph{open}.

\begin{example}\label{ex:running}
    In this example, we build a justification frame to express the transitive closure of a graph.
    So let $V$ be a set of nodes. Define the open elements to be the set of elements $\Edg(v,w)$ and $\tild\Edg(v,w)$ with $v, w \in V$ and the defined elements to be the set of elements $\Path(v,w)$ and $\tild\Path(v,w)$ with $v, w \in V$. The facts encoding the edges of a graph are open. This means that they can freely change and thus act as parameters, whereas the defined facts are constrained by the rules seen below. We define the rules\footnote{Normally in justification theory you add rules dual to these as well; rules that define $\tild\Path(v,w)$ and are compatible with the rules for $\Path(v,w)$.} for $\Path(v,w)$:
    \begin{align*}
        &\Path(v,w) \gets \Edg(v,w); \\
        &\Path(v,w) \gets \Path(v,x), \Path(x,w)
    \end{align*}
    for all $v,w,x \in V$. This encodes that $\Path$ is the transitive closure of $\Edg$.
\end{example}

A \emph{justification} over a justification frame is a subset of $\rules$ containing at most one rule for each defined fact. We can also view a justification as a partial function from $\F$ to subsets of $\F$ by mapping a defined fact to the body of the rule in the justification if it exists. Differently, a justification can also be viewed as a graph, where we have arrows going from the head to elements in the body.

A justification is \emph{locally complete} if for each defined fact occurring in the body of a rule in the justification, there is a rule with that fact as the head in the justification. If you view the justification as a graph, it means that no complete path ends in an open fact.

\begin{example}\label{ex:running_justification}
    Let $V=\setl{a,b,c}$ in the setting of Example~\ref{ex:running}. Suppose that $\Edg(a,b)$ and $\Edg(b,c)$ hold. The following locally complete justification gives an explanation why $\Path(a,c)$ holds.
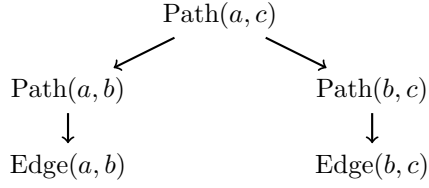
    \begin{center}
        \begin{tikzpicture}[transform shape]
        \node (a) at (2, 2) {$\Path(a,c)$};
        \node (b) at (0, 1) {$\Path(a,b)$};
        \node (c) at (4, 1) {$\Path(b,c)$};
        \node (d) at (0, 0) {$\Edg(a,b)$};
        \node (e) at (4, 0) {$\Edg(b,c)$};
        \tikzstyle{EdgeStyle}=[style={->}]
        \Edge(a)(b)
        \Edge(a)(c)
        \Edge(b)(d)
        \Edge(c)(e)
        \end{tikzpicture}
    \end{center}
\end{example}

A \emph{branch} in a justification frame is either a finite sequence of defined facts together with an open fact; or an infinite sequence of defined facts. In Example~\ref{ex:running_justification}, the following is a branch of the illustrated justification
\begin{equation*}
\Path(a,c) \rightarrow \Path(a,b) \rightarrow \Edg(a,b).
\end{equation*}

A \emph{branch evaluation} is then a mapping from branches to facts. The \emph{well-founded branch evaluation} for example maps finite branches to its open fact and infinite branches to $\Tr$ if it has a negative tail, to $\Fa$ if it has a positive tail, and to $\Un$ otherwise. The \emph{stable branch evaluation} maps to the first element that has a different sign as the first element if it exists, otherwise it maps positive loops to $\Fa$ and negative loops to $\Tr$ and finite branches of the same sign to the last element.

\begin{example}
Take the following logic program
\begin{align*}
& p \gets \tild p
\end{align*}

The only locally complete justification for $p$ is given below
\begin{center}
    \begin{tikzpicture}[transform shape]
    \node (a) at (0, 1) {$p$};
    \node (b) at (0, 0) {$\tild p$};
    \tikzstyle{EdgeStyle}=[bend left, style={->}]
    \Edge(a)(b)
    \Edge(b)(a)
    \end{tikzpicture}
\end{center}
The well-founded branch evaluation maps the branch $p \rightarrow \tild p \rightarrow \cdots$ to $\Un$, while the stable branch evaluation maps it to $\tild p$.
\end{example}

A fact is \emph{supported} by a locally complete justification in a subset $I \subseteq \F$ if each branch in the justification starting with that fact is mapped to $I$ under the branch evaluation. A fact is \emph{supported} in $I$ if there is some locally complete justification supporting that fact in $I$.

\begin{example}
    Let the setting be the same as in Example~\ref{ex:running_justification}. Take the set $I=\setl{\Edg(a,b), \Edg(b,c)}$. The fact $\Path(a,c)$ is supported in $I$ under the
    Well-founded branch evaluation, a justification supporting $\Path(a,c)$ is already given in Example~\ref{ex:running_justification}.
\end{example}

The \emph{support operator} maps a subset $I$ to the facts it supports. Intuitively, you can view it in the following way: $I$ is the set of what you already know, and the support operator maps $I$ to the facts that can be derived when the facts in $I$ holds. To make the support operator iterable, we just add the elements of $I$ back to the output of the support operator. This gives us the \emph{extended support operator}. The (least) fixed points hereof are used to define justification semantics. In \cite{lpnmr/DeneckerBS15}, it is proven that the justification semantics of the stable and well-founded branch evaluations match their equally-named logic programming semantics.

\section{Central objective}

As mentioned in the background, the central objective is to research extensions justification theory in several directions, to ultimately support legal decision systems.

\subsection{First-order justifications}

One limitation of the current state of justification theory is that it only deals with \emph{ground} rules. This limitation is noticeable in lazy grounding techniques since ground justifications are much larger than first-order justifications, see e.g.,~\cite{ijcai/BogaertsW18}. Therefore, we see it fit to extend the theory to a first-order formalism. Keeping track of variables in loops makes it challenging to devise branch evaluations corresponding to the stable and well-founded semantics. In \cite{corr/MarpleSG17}, the authors compute stable models without grounding with a goal-directed method. They use techniques that resemble ones in justification theory, but they work in a first-order setting. Therefore this work provides a base that we can build on.

\subsection{Extending logic programing}
In \cite{lpnmr/DeneckerBS15} it is mentioned that nesting of justification frames can be used to define semantics for logic programming with aggregates.
One research question to tackle is whether this idea can be generalized to add arbitrary new logical constructs to logic programming.
In the introduction, we discussed the use for modal operators such as alethic and deontic modalities. Therefore extending logic programming with these modalities is of particular interest to me. Some work has been done already in this area, most remarkably MOLOG, a system that extends Prolog with modal logic, see \cite{ngc/Cerro86}.

\subsection{Nondefective justification frameworks}
One advantage of justification theory is that both positive and negative facts can occur in the head of a rule. However this brings a caveat with it: a fact and its negation should not be supported simultaneously; if for example $p$ and $\tild p$ are supported, then there should be inconsistencies in the rules. If this is not the case, we call the branch evaluation \emph{defective}. Stated differently, a branch evaluation is non-defective if the following holds: if the rules are consistent, then for each non-logical fact $p$, we have that at most one of $p$ and $\tild p$ is supported.

In recent work \cite{nmr/MarynissenPBD18}, we proved that this is the case for the well-founded, stable, Kripke-Kleene and Clarke completion branch evaluations. The proofs for these results are not easily generalisable to other branch evaluations. If we extend justification theory to a first-order setting or extend it with modal operators, we still need to prove that the corresponding branch evaluations are not defective. Therefore, we want to investigate which properties a branch evaluation needs to be non-defective to obtain general classes of non-defective branch evaluations.

\subsection{Possible practical applications}
As mentioned in the introduction, the research stipulated here can be used in more practical applications. A type of application that interests me, in particular, is an interactive decision enactment system. This is a system that makes a legal decision, such as granting an environmental permit, with interaction from the user. The interaction is in the form of the user providing partial information either from input or output.

Central to such a decision system is a modelling of a particular legislative text into a logical format. With this modelling, one can perform various kinds of reasoning. An important example of a type of reasoning in a legal context is providing explanations; explaining why the system has inferred some information. Since justifications can be seen as explanations, it is natural to consider a system based on justification theory. However, the underlying framework should be extended in order to support logical constructs present in legislation, e.g., deontic and alethic operators.

In such a system it often occurs that if you are in a particular partial configuration of the decision, some information is not needed any more, the information becomes \emph{irrelevant}. The system should then only query relevant information to improve usability. In \cite{icsc/DeryckDMV19}, we developed a notion of relevance in terms of justification theory.

Since both explanations and relevance can be defined with justification theory, we believe that justification theory is a good candidate for the back-end for an interactive decision enactment system.

More practically, I plan to work on Belgian law regarding termination of employee contracts. An example decision could be deciding how much compensation the employee receives.

\subsection{Expressing new semantics}
While many semantics of logic programming are captured by justification theory, some others cannot be expressed yet.
For example, how do we express the ultimate stable and well-founded semantics \cite{iclp/DeneckerPB01}? Ultimate stable semantics is used to define semantics for logic programs with aggregates, see~\cite{iclp/DeneckerPB01}.

One idea is to alter the rules of the justification frame and use the support operator on this revised justification frame to define models by means of fixed points. If this approach is suitably general, it allows us to define ultimate semantics for various other formalisms.

Another idea is to devise a new branch evaluation for ultimate stable semantics and ultimate well-founded semantics. Several questions arise: how does the ultimate branch evaluations relate to the original branch evaluations? Can we define an `ultimate' operator on branch evaluations that maps a branch evaluation to its corresponding ultimate branch evaluation?

When new semantics are defined they are often given terms of fixed points of a particular operator. Since justification semantics are also given in terms of fixed points of an operator associated with a branch evaluation, we can wonder if there is a systematic way to determine the branch evaluation corresponding to the original semantics?

\bibliographystyle{splncs04}
\bibliography{lib}

\end{document}